\documentclass{optica-article}

\journal{opticajournal} % for journals or Optica Open

\articletype{Research Article}

\usepackage{graphicx}% Include figure files
\usepackage{dcolumn}% Align table columns on decimal point
\usepackage{bm}% bold math
\usepackage{braket}
\usepackage[english]{babel} %To allow zotero to not throw errors 
\usepackage{xcolor}
\UseRawInputEncoding

\begin{document}

\title{Phase matching in Vector Beam Driven High Harmonic Generation with 3D-printed Gas Cells}

\author{D. Attiyah\authormark{1,*}, H. Allison\authormark{1}, J. To\authormark{1}, P. Kazansky\authormark{2}, D. Schmidt\authormark{3}, C. Gardner\authormark{1}, V. Flores\authormark{1}, J. Lewis\authormark{1}, C.G. Durfee\authormark{3}, and F. Dollar\authormark{1}}

\address{\authormark{1}STROBE, NSF Science \& Technology Center, University of California, Irvine\\
\authormark{2}Optoelectronics Research Centre, University of Southampton\\
\authormark{3}Department of Physics Colorado School of Mines}

\email{\authormark{*}dattiyah@uci.edu} %% email address is required; see note below about the corresponding author designation

% use {asbstract*} to suppress the copyright line. Copyright information will be added in production

\begin{abstract*} 
We present experimental results of high harmonic generation(HHG) driven by a 1300 nm beam in three different polarization states: linear, radial, and azimuthal. We found that the optimal pressure for phasematching was roughly twice as high for the vector beam drivers than the linear driver. We attribute this difference in pressure primarily to the Gouy phase, which differs by a factor of two between the linear and vector polarization states. We demonstrate a target for HHG that produces a uniform pressure profile in the interaction region that is nearly identical to the backing pressure, and preserves the mode of the driving beam. We provide characterization and validation of this technique through flow simulations and experimental measurements.
\end{abstract*}

%%%%%%%%%%%%%%%%%%%%%%%%%%  body  %%%%%%%%%%%%%%%%%%%%%%%%%%
\section{Introduction}
Vector beams, or laser beams with spatially dependent polarization, have garnered interest across many disciplines in recent years. A common example is the radially polarized beam, which is locally linearly polarized at each point in space, but with a macroscopic polarization state that points away from the beam center. At the center, the superposition of these varying polarization vectors creates a singularity, resulting in a characteristic ring-shaped intensity profile. The orthogonality and tight focusing capabilities \cite{dorn_sharper_2003} of these vector beams have enabled advances in optical communications \cite{milione_4_2015,milione_using_2015,zhao_high-base_2015,zhu_compensation-free_2021}, microscopy \cite{torok_use_2004,chen_imaging_2013,rong_super-resolution_2015}, and optical tweezers \cite{zhan_trapping_2004,yan_radiation_2007, michihata_measurement_2009,kozawa_optical_2010, bhebhe_vector_2018}. These same properties have also expanded the capabilities of ultrafast lasers, or lasers with Full-width at Half-Maximum (FWHM) pulse durations of $<100$ femtoseconds ($10^{-15}$ seconds), in applications such as materials processing \cite{hnatovsky_revealing_2011,jin_dynamic_2013,ouyang_tailored_2015,skoulas_biomimetic_2017}, magnetic probing for materials research \cite{blanco_ultraintense_2019,fujita_nonequilibrium_2018} and laser driven particle acceleration \cite{salamin_direct_2008,marceau_femtosecond_2013}. 

High Harmonic Generation (HHG) is a nonlinear, frequency upconverting process that produces harmonic radiation, or light with wavelengths at integer multiples of the driving beam. When driven by near-infrared (NIR) ultrafast beams, this interaction can produce light in the extreme ultraviolet (EUV) wavelength range with pulse durations on the attosecond ($10^{-18}$ seconds) scale \cite{lhuillier_multiply_1983,hentschel_attosecond_2001,paul_observation_2001}. The microscopic physics of HHG is often described by a semi-classical three-step recombination model \cite{corkum_plasma_1993} consisting of tunnel ionization of the electron, acceleration by the laser field, and recombination with the parent atom accompanied by the emission of the upconverted photon \cite{lewenstein_theory_1994}. When the driving laser has a longer wavelength, the electron is accelerated for a longer period of time, causing it to gain more energy and release a higher energy photon upon recombination. Therefore, driving beams with longer wavelengths are capable of producing higher energy, or shorter wavelength, photons \cite{shan_dramatic_2001,tate_scaling_2007}. The emitted spectrum from a single atom would appear as a plateau across the relevant wavelengths.

Phasematching describes the macroscopic coherent buildup of harmonic radiation in the interacting medium by minimization of the phase mismatch between the driving beam and the harmonic beam. The flux of the harmonic beam acquires a Gaussian-like shape and can be significantly increased when phasematching conditions are met \cite{rundquist_phase-matched_1998}. The total phase mismatch between the driving field and a given harmonic order can be written as a sum of the dominant sources of dispersion \cite{durfee_phase_1999}: $\Delta k = \Delta k_{\textrm{plasma}} + \Delta k_{\textrm{neutral}} + \Delta k_{\textrm{Gouy}} + \Delta k_{\textrm{intrinsic}}$
where the right hand side represents the phase mismatch due to propagation in plasma and neutral gas \cite{lhuillier_propagation_1990}, focusing geometry \cite{balcou_phase-matching_1993}, and the intrinsic nonlinear phase \cite{salieres_coherence_1995,peatross_intensity-dependent_1995,lewenstein_phase_1995,balcou_generalized_1997}, respectively. These terms depend on target parameters such as ionization potential and density profile, as well as laser properties including intensity, wavelength, polarization, and mode \cite{lhuillier_theoretical_1991,lhuillier_high-order_1993,platonenko_generation_1998}. Phase matching at longer wavelengths has been demonstrated across the mid-infrared wavelength region, where higher densities were used to reach suitable phase matching conditions \cite{popmintchev_extended_2008,popmintchev_phase_2009,chen_bright_2010,popmintchev_bright_2012}.

\begin{figure*}[htbp]
    \centering
    \includegraphics[width=1\linewidth,trim={0cm 3cm 0cm 3cm},clip]{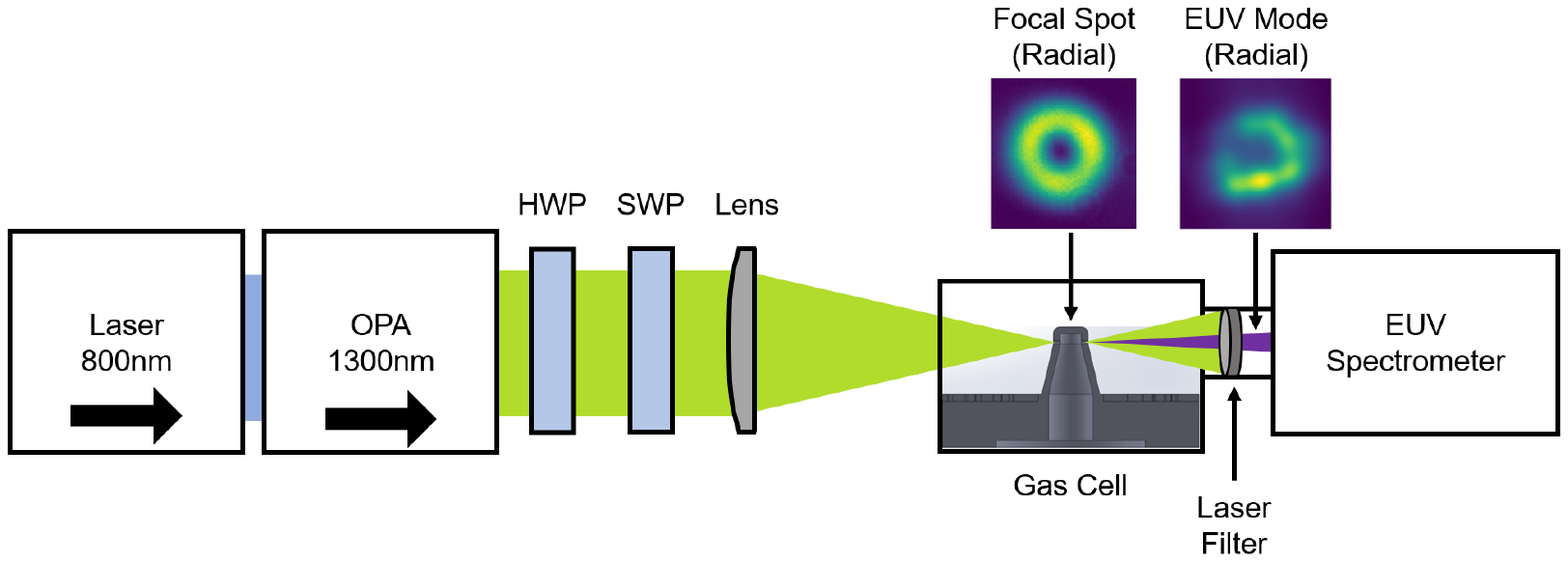}
    \caption{Experimental setup showing the "S" waveplate that creates the vector polarization states which are then focused into the gas cell, where high harmonics are generated. The laser light is filtered out so that the harmonics can be measured with the EUV spectrometer.}
    \label{fig:exp_setup}
\end{figure*}

There has been growing interest in studying HHG driven by vector \cite{hernandez-garcia_extreme_2017,kong_vectorizing_2019}, vortex \cite{zurch_strong-field_2012,gariepy_creating_2014,kong_controlling_2017}, vector-vortex \cite{de_las_heras_extreme-ultraviolet_2022} and circularly polarized beams \cite{hickstein_non-collinear_2015,huang_polarization_2018,dorney_controlling_2019}. To produce high-brightness harmonic radiation with complex polarization states, it is critical to understand the phasematching conditions, especially when using longer wavelength driving beams to reach shorter wavelengths. However, polarization studies to date have largely been limited to NIR and optical wavelengths.

In this work, we present experimental results of HHG using a 1300 nanometer (nm) driving beam in three different polarization states: linear, radial, and azimuthal. We found that the optimal phasematching pressure differed by a factor of two between the linear and vector polarized driving beams. We attribute the difference in pressure to the Guoy phase term in the phasematching equation, which is two times larger for the vector beam than it is for the linear beam. We use 3D-printed gas cells designed to optimize the pressure profile in the interaction. We provide characterization and validation of this technique through flow simulations and experimental results.

\section{Methods}

%Methods
The experiment was performed with the output of a optical parametric amplifier (OPA, Light Conversion TOPAS-Prime) driven by a Ti:Sapphire laser system (Spectra Physics Solstice ACE). With a central wavelength of 1300 nm, a pulse duration of 35 femtoseconds (fs) and a pulse energy up to 1 milliJoule (mJ), the linearly polarized beam was focused to intensities of $1.9\pm 0.4\times 10^{14}$ $\textrm{W/cm}^{2}$ using an anti-reflection (AR) coated lens with a focal length of 200 millimeters (mm). The focal spot size was inferred from a measurement of the second harmonic signal focal spot. With a beam radius of $w_0 =37.8 \pm 3$ micrometers ($\mu m$), the Rayleigh length was calculated to be $z_R=3.3 \pm 0.3 \textrm{mm}$. The experimental configuration to create the vector beams was identical, except for a half-waveplate (HWP) and "S-waveplate" (SWP) \cite{beresna_radially_2011} that were placed before the lens. The HWP was used to set the angle of the incident polarization angle relative to the axis of the SWP. This determines if the output polarization state is radial, azimuthal, or a superposition of the two. In the configuration with the SWP, the energy of the input beam was increased by a factor of 2.79 to maintain the same peak intensity as the linearly polarized beam. A simplified diagram of the experimental setup is shown in Figure \ref{fig:exp_setup}.

The laser was focused into a 3D-printed gas cell filled with Argon, designed to provide a uniform density profile. The cell was fabricated from a Stratasys Objet260 Connex3 machine using VeroClear material. The entry and exit orifices of the cell are $500\mu m$ in diameter, and the length of the interaction region was 2.5 mm. The gas cell was positioned at the focal plane of the laser with an estimated uncertainty of $\pm 1mm$ due to manual alignment of the visible plasma channel. This corresponds to $\sim 30\%$ of the Rayleigh length, where the on-axis intensity remains within $10\%$ of its peak value. Small adjustments to the gas cell position (within $100 \mu m$) optimized the peak of a single harmonic and increased the overall flux of the measured spectrum. Larger steps decreased the overall flux and shifted the center of the spectrum. The backing pressure was varied during the experiment between 50 and 550 torr. Differential pumping isolated the flow out of the gas cell from the remaining vacuum system.  

The output radiation produced in the gas cell was characterized in both its spectral content and spatial profile. The harmonic radiation was isolated from the fundamental driving wavelengths using ultrathin x-ray filters made of Aluminum, Zirconium, or a combination of the two depending on the diagnostic used. These materials are opaque to the fundamental wavelength but transmit EUV wavelengths. The harmonic spectra were measured with a home-built flat-field spectrometer \cite{neely_multi-channel_1998}, which used a variable line spacing imaging grating (Hitachi, 1200 lines/mm) \cite{kita_mechanically_1983}. Time-integrated spectra were measured \textit{in situ} with an x-ray CCD (Andor-Newton). The spatial profile of the harmonic radiation was measured by placing the x-ray CCD directly in the filtered beam path, rather than at the output of the spectrometer.

\section{Results}
\begin{figure} [htbp]
    \centering
    \includegraphics[scale=0.38]{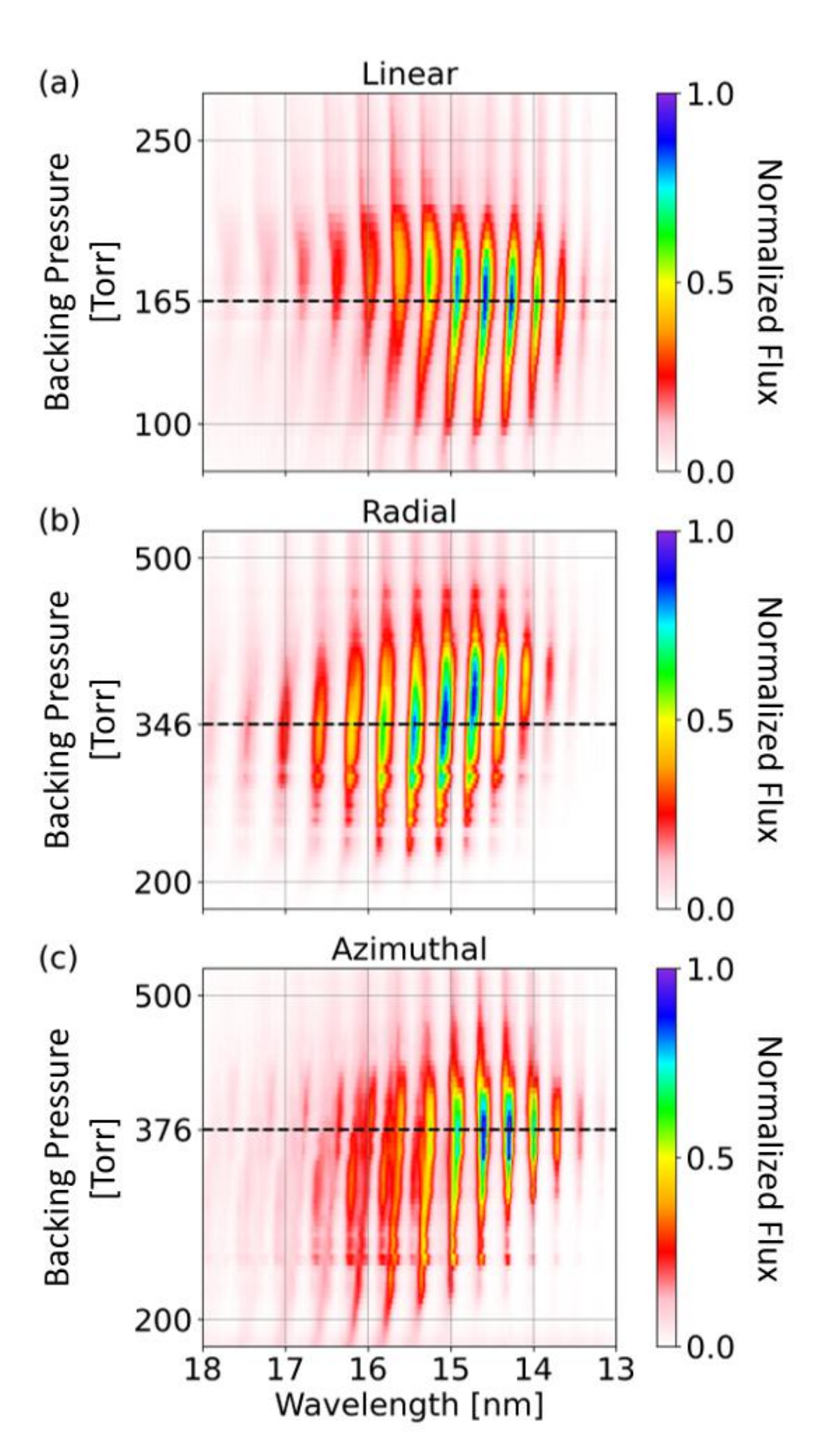}
    \caption{EUV spectra measured at individual pressures, then binned, stacked, and interpolated to plot on linear axes. Pressure scans were performed with a 1300 nm driving beam of linear (top row), radial (middle row), and azimuthal (bottom row) polarizations. Dashed lines mark the pressure at the peak flux. Spectra were taken with Zirconium filters with a total thickness of 400 nm.}
    \label{fig: Pressure_Scans}
\end{figure}

The harmonic radiation through 400 nm of Zirconium is plotted as a function of pressure in Fig. \ref{fig: Pressure_Scans}). For each of the polarization states measured, there is a well-defined maxima in the normalized flux that can be observed in both the wavelength and the backing pressure. Considering only the spectral axis, the signal decays (rather than plateauing) on both sides of this peak for the same wavelengths for all polarization states. The spectral peak is at $\sim14.6$ nm, and remained relatively constant as the pressure was varied. Along the pressure axis, there is a similar peak and decay in the normalized flux. The pressure peak for the radial polarization case (346 torr) is approximately the same as the peak for the azimuthal polarization case (376 torr), both of which are roughly twice the peak found for the linear polarization case (165 torr).

\begin{figure} [htbp]
    \centering
    %[trim={left bottom right top},clip]
    \includegraphics[scale=0.25]{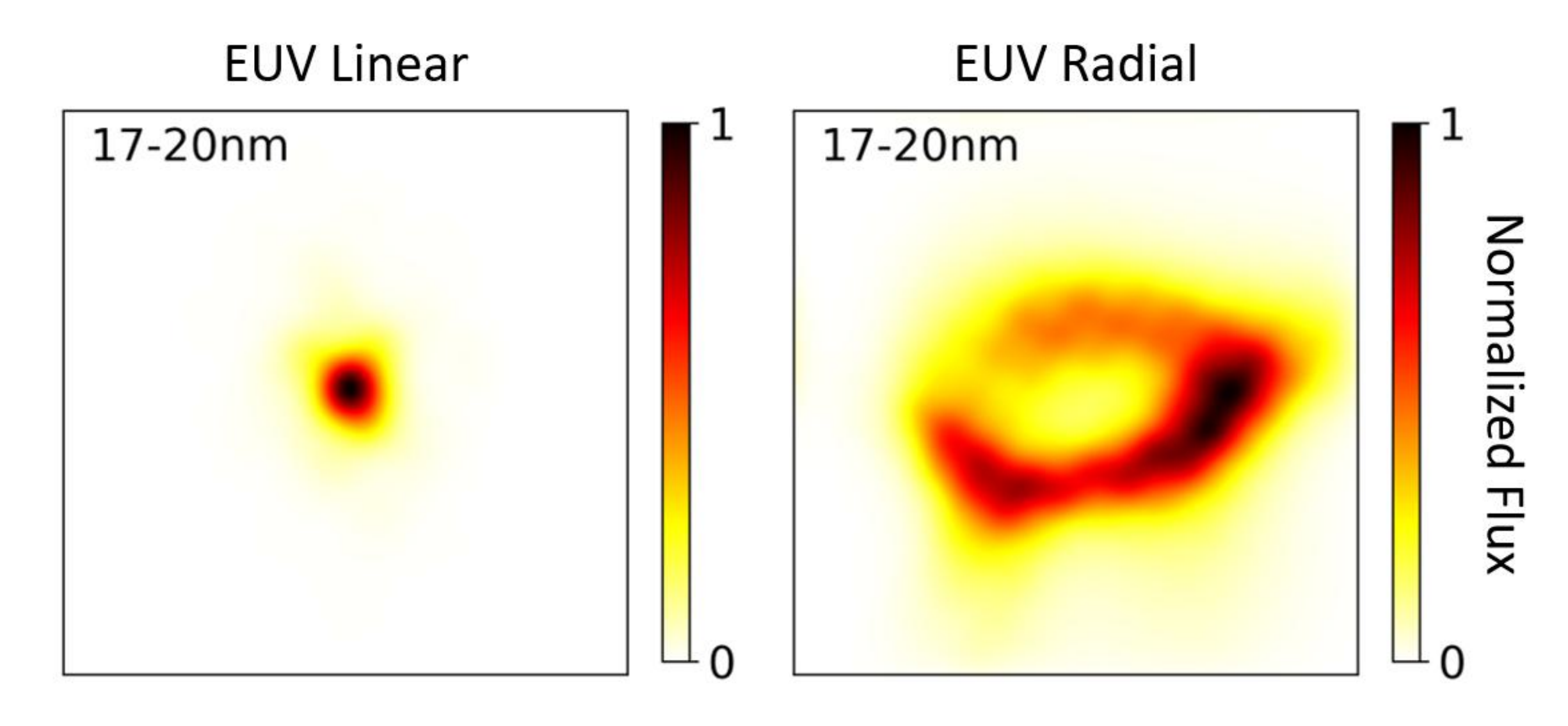}
    \caption{Direct measurements of the EUV light produced by both the linear and vector polarized driving beam. High frequency components of the measured signal, such as the mesh grid of the ultrathin filters, were removed using a Fourier filter.}
    \label{fig: Direct EUV}
\end{figure}

The entire spatial profile of the harmonic radiation was captured with the x-ray CCD (See Fig. \ref{fig: Direct EUV}). In order to isolate a narrow spectral window, both aluminum and zirconium filters were used to provide a bandpass of 17 to 20 nm. The linearly polarized driving beam generated harmonic radiation with a Gaussian spatial profile, whereas for the vector beams the generated harmonic radiation was ring shaped. Example profiles for the linear and radial polarization drivers are shown in Fig. \ref{fig: Direct EUV}. The harmonic flux is comparable between the two polarization states.

\section{Discussion}

The peak in the pressure scan for each polarization state indicates phasematching conditions are being met. We can write the phasematching equation as:
\begin{eqnarray}
    &\Delta k \approx P \bigg[\underbrace{ - \frac{\eta \lambda_0 r_e}{k_BT}\left(q-\frac{1}{q}\right)}_{\Delta k_{\textrm{plasma}}} + \underbrace{\frac{2 \pi q(1-\eta)\Delta n}{\lambda_0 k_B TN_{1atm}}}_{\Delta k_{\textrm{neutral}}} \bigg] \underbrace{- \frac{q \pi D^2}{4 \lambda_0 f^2}}_{\Delta k_{\textrm{Gouy}}} \label{eq:PM_pressure}
\end{eqnarray}

where the gas pressure, $P$, has been factored out of the plasma and neutral dispersion terms, $\eta$ is the fractional ionization level, $\lambda_0$ is the laser wavelength, $r_e$ is the classical electron radius, $k_B$ is the Boltzmann constant, $T$ is the temperature, $q$ is the harmonic order, $\Delta n$ is the difference in index of refraction between the fundamental beam and the harmonic beam, $N_{1atm}$ is the gas density at atmospheric pressure, $D$ is the beam diameter, $f$ is the focal length, and the intrinsic phase has been neglected. This assumption is valid when the target is near focus ($z=0$) in long focal length geometries that satisfy the paraxial approximation. The divergence half-angle in our geometry is $\theta= \arctan(D/2f)= 0.63 ^{\circ}$, which satisfies this condition. The same approach was used in studies performed in similar pressure ranges \cite{popmintchev_extended_2008, popmintchev_phase_2009}. We are also assuming a free focusing geometry such that the phase mismatch due to geometry may be described with the Gouy phase. If all variables are fixed and only the pressure is changed, $\Delta k$ varies linearly and intercepts 0 when the expression in the bracket on the right hand side of Equation \ref{eq:PM_pressure} is equal to the Gouy phase term. This pressure-tuning phasematching \cite{durfee_phase_1999, rundquist_phase-matched_1998} resulted in the peak at 165 torr for the linear polarization state (Figure \ref{fig: Pressure_Scans} (a)). The peak along the spectral axis at 14.6 nm is also an indication of phasematching. The optimal phasematching pressure and wavelength measured with the linear polarization state is consistent with previous phasematching studies with 1300 nm driving beams \cite{popmintchev_extended_2008}.

The vector polarization states produce similar peaks along each axis (Figure \ref{fig: Pressure_Scans} (b) and (c)), indicating that phasematching conditions are being met in these cases as well. Similar to the linear polarization state, the peak along the spectral axis for both vector polarization states occurs at, or near, 14.6 nm. The location of this peak in spectral space is determined by the wavelength and intensity of the driving beam \cite{corkum_plasma_1993,lewenstein_theory_1994}. The wavelength was 1300nm for all polarization states; therefore, the peak intensity must have been similar for each case. This can be further supported by a model of the intensity profile for each polarization state in the same focusing geometry. The radially polarized electric field amplitude can be expressed as a combination of Hermite-Gaussian modes: $\ket{u_{\textrm{Rad}}}=\ket{HG_{10}}\hat{x}+\ket{HG_{01}}\hat{y}$. Similarly, the azimuthally polarized field amplitude can be written as $\ket{u_{\textrm{Az}}}=\ket{HG_{01}}\hat{x}-\ket{HG_{10}}\hat{y}$. The intensity profile of these two ideal vector beam states are identical. By comparing these profiles to a linearly polarized, Gaussian profile, we find that the peak intensity of the linear state is $\sim2.7$ times greater than the peak intensity of the vector beams. In our experiment, we increased the energy of the vector beams by a factor of 2.79, which agrees with this model.

The observed shift in the optimal phasematching pressure between the linear and vector polarization states is likely primarily due to the difference in the Gouy phase, $\Phi_{\textrm{Gouy}}$. For a linearly polarized, Gaussian beam in a free-focusing geometry, this term varies as $-\textrm{arctan}(z/z_R)$, where $z_R$ is the Rayleigh length. Near $z=0$, we can make the approximation $\Phi_{\textrm{Gouy}} = k_{\textrm{Gouy}} z \approx -z/z_R$. The phase mismatch, $\Delta k_{\textrm{Gouy}}= q k_{\textrm{Gouy}, \omega_0} - k_{\textrm{Gouy}, q\omega_0}$, is dominated by the fundamental term: $q k_{\textrm{Gouy}, \omega_0}=-q/z_R=-q\lambda_0/\pi w_0^2$. Plugging in the expression for the beam waist, $w_0=2\lambda_0f/\pi D$, yields the expression for the Gouy phase term in Equation \ref{eq:PM_pressure}. By modeling the vector beams as combinations of $HG_{10}$ and $HG_{01}$ modes, we find that the Gouy phase for both the radial and azimuthal polarization states is $-2\textrm{arctan}(z/z_R)$. This doubling of the Gouy phase was experimentally verified for radially polarized beams in 2016 by Kaltenecker, \textit{et al.} \cite{kaltenecker_gouy_2016}. Equation \ref{eq:PM_pressure} describes the phase mismatch for the linear polarization state. We can write out the equivalent equation for the vector polarization states and set each equal to zero. We can solve for the optimal phasematching pressure of the vector polarization states, $P_{\textrm{VP}}$, in terms of that for the linear polarization state, $P_{\textrm{LP}}$. Because the intensity is the same for all cases, the expression in the brackets is unchanged. Since the relationship between the Gouy phase terms is $\Delta k_{\textrm{Gouy,VP}} = 2\Delta k_{\textrm{Gouy,LP}}$, then, in the approximation that the intrinsic phase is negligible, the relationship between the pressures must be $P_{VP} \approx 2P_{LP}$. The measured relationship between $P_{VP}$ and $P_{LP}$ at $\sim 14.6$ nm was $\sim2.1$ for the radial polarization state and $\sim2.3$ for azimuthal polarization state, in good agreement with our model.

The observed discrepancies may be due, in part, to the intrinsic phase, which varies as the gradient of the intensity \cite{salieres_coherence_1995,peatross_intensity-dependent_1995,lewenstein_phase_1995,balcou_generalized_1997}, and may impact the phasematching in the longitudinal direction. When the target is at the focal plane ($z=0$ mm), the gradient of the intensity in the longitudinal direction is 0. However, the error bar in our focal plane alignment was $\pm 1mm$. At $z=1$ mm , the peak intensity of a linearly polarized, Gaussian beam decreases by $\sim 10\%$ (assuming a focal geometry identical to the one used in our experiment). The same is true for the vector polarization states modeled by HG modes. This could result in the longitudinal intrinsic phase becoming significant enough to cause the discrepancies observed in the phasematching pressure ratios. 

The intrinsic phase and other intensity fluctuations may also be responsible for the different trends in the slope of the phasematching curves in Figure \ref{fig: Pressure_Scans}. The linear polarization state produces a curve with a negative slope, whereas the radial and azimuthal polarization states produce curves with positive slopes (of different magnitudes). It is possible that imperfections in the driving beam intensity profile could cause variations in the intrinsic phase for a particular harmonic order, which could change the optimal phasematching pressure for that order. These imperfections could have been introduced by the nonlinear processes in the OPA. Alternatively, it is possible that the S-waveplate could introduce polarization or mode defects if the incidence angle is not perfectly normal. Theoretically, the radial and azimuthal polarization states should be identical, but asymmetry in either beam profile could have caused the difference in slope between the two states. 

The direct measurement of the harmonic radiation produced by the radially polarized beam (Figure \ref{fig: Direct EUV} (b)) confirms that the ring-shaped mode is preserved in the interaction. While previous studies have shown that vector-polarized driving beams generate vector-polarized harmonic radiation \cite{hernandez-garcia_extreme_2017,kong_vectorizing_2019,de_las_heras_extreme-ultraviolet_2022}, they were performed at shorter wavelengths, limiting the harmonic order q to a maximum of $\sim27$. With longer wavelength driving beams, both the maximum energy and the harmonic order q increase substantially. For instance, in the 17--20 nm spectral range transmitted by the x-ray filters in our experiment, harmonic orders from the 65th to the 75th were measured. At such high orders, the divergence of harmonic radiation is nearly identical between orders, leading to spatial overlap. The observation of a well-defined ring structure at these high harmonic orders indicates that the vector polarization state is preserved in the interaction. As the harmonic order increases, the impact from mode imperfections also increases, as the interference between each mode accumulates. Theses imperfections could contribute to the transverse intrinsic phase, which could affect the divergence and structure of the EUV profile. Improving the driving beam mode quality would likely increase conversion efficiency and harmonic mode quality.

Our phasematching measurements were enabled by the 3D-printed gas cell. The gas cell provides a cost-effective, customizable alternative to simple gas jets. It consists of a gas-filled interaction region with entrance and exit orifices for the laser to pass through. The free-focusing geometry, similar to a gas jet, enabled the vector polarization states to propagate freely through the interaction region. Because there is no optical guiding, the mode is preserved in the interaction, enabling us to create harmonic radiation with vector polarization states.

\begin{figure}[htbp]
    \centering
    \includegraphics[scale=0.3]{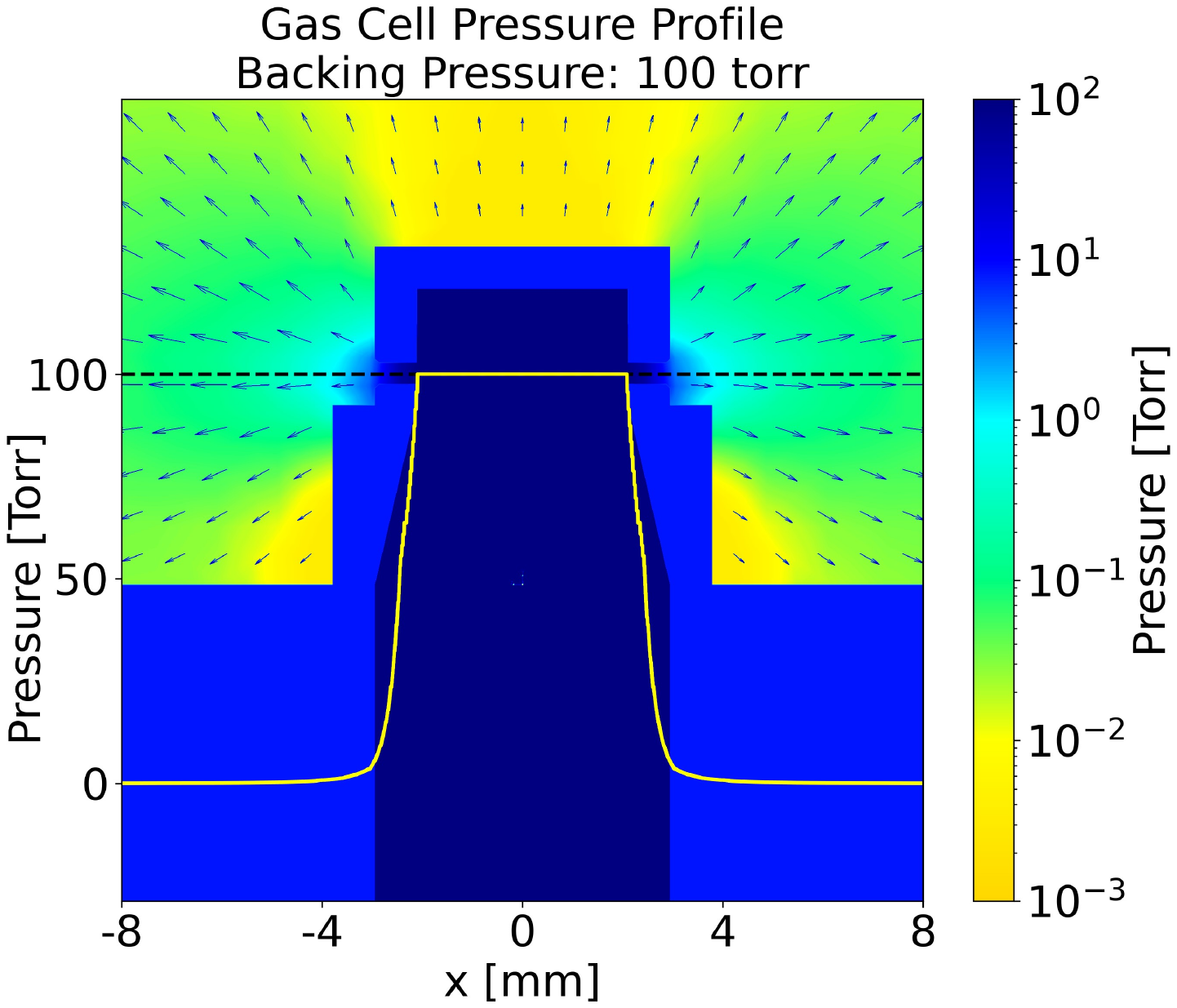}
    \caption{A flow simulation of the gas cell shows an interaction pressure nearly identical to the backing pressure of 100 Torr. The color bar (log scale) indicates the pressure in the image. The arrows in the image represent the direction of the gas flow. A lineout along the dotted black line is overlaid on the image (yellow).}
    \label{fig:cell_char}
\end{figure}

Our phasematching model assumes that the gas cell creates a confined region of uniform pressure, an assumption we validate through numerical simulations and experimental measurements. We modeled the gas density profile using SolidWorks Flow Simulation, a CAD-embedded Computational Fluid Dynamics (CFD) tool that solves the complete Navier-Stokes equations. The results show that the pressure remains very uniform (within $0.05\%$) throughout the interaction region and drops off rapidly within the orifices (see Figure \ref{fig:cell_char}). The simulations also show that a backing pressure of 100 Torr yields a nearly identical interaction pressure of 99.73 Torr. Using the gas cell and a linearly polarized 800 nm driving beam, we experimentally measured an optimal phasematching pressure of 70 torr with peak flux at a wavelength of 27.6 nm, which agrees with previous studies \cite{chen_thesis_2012}. 

Using the 3D-printed gas cell as the target in our experiment was an integral component of obtaining these results. High pressure gas jets inherently produce a non-uniform density profile, making them less ideal for phase matching studies. Additionally, the vacuum system can only accommodate a certain amount of gas flow. The use of gas cells to confine the gas enables much lower flows than a gas jet. Even though the optimal phasematching pressure for the vector polarization states was roughly twice that of the linear polarization state, this regime was accessible because of the gas cell. Using 3D-printed gas cells rather than traditional gas cells enabled rapid prototyping to optimize the measured HHG flux. Design flexibility was limited with the traditional gas cells available, making it difficult to generate harmonic spectra with a high signal-to-noise ratio.

\section{Conclusion}
We have shown experimental results of phasematched HHG in a 3D-printed gas cell using a 1300 nm driving beam in linear and vector polarization states. The spatial profiles of the EUV light match those of the corresponding driving beams, confirming spatial mode transfer. Maintaining the same experimental parameters for each polarization state and varying only the backing pressure, we found that the optimal phasematching pressure was roughly two times higher for the vector polarization states than the linear state. We believe that this is primarily a consequence of the spatial mode of the vector beams that doubles the Gouy phase, and not a polarization dependent phenomenon. This work was enabled by 3D-printed gas cells that create a confined region of uniform pressure conducive to phasematching. The phasematching conditions for the vector polarization states were achievable because the gas cell preserved the mode of the driving beam and provided an interaction pressure nearly identical to the backing pressure.

Our findings suggest that it may be useful to further explore phasematching conditions for vortex driving beams in HHG. While these beams carry a unique phase structure, the ring-shaped intensity profile is very similar to the vector beams used in our experiment. The 3D-printed gas cell facilitates HHG with such unconventional spatial modes and supports alternative geometries such as noncollinear setups and variable length designs. The low cost and design flexibility of the gas cells provide exciting opportunities for experimental prototyping and customized targetry in HHG.

\begin{backmatter}
\bmsection{Funding}
This work was supported by STROBE: A National Science Foundation Science \& Technology Center under Grant No. DMR-1548924. This work was also supported by the National Science Foundation under Award No. PHY-1753165 and PHY-1903709.

\bmsection{Disclosures}
\noindent The authors declare no conflicts of interest.

\bmsection{Data availability} Data underlying the results presented in this paper are not publicly available at this time but may be obtained from the authors upon reasonable request.

\end{backmatter}

%%%%%%%%%%%%%%%%%%%%%%% References %%%%%%%%%%%%%%%%%%%%%%%%%

%%%%%%%%%% If using BibTeX:
\bibliography{mbib}

\end{document}